%% file: main.tex
\documentclass[lettersize,journal]{IEEEtran}
\usepackage[utf8]{inputenc}
\usepackage[cmex10]{amsmath}
\usepackage{amsbsy,pifont,verbatim,amsfonts,amssymb}
\usepackage{array}
\usepackage[caption=false,font=normalsize,labelfont=sf,textfont=sf]{subfig}
\usepackage{textcomp}
\usepackage{stfloats}
\usepackage{url}
\usepackage{verbatim}
\usepackage{graphicx}
\usepackage{algorithm2e}
\usepackage{hyperref}
\usepackage{xcolor}
\usepackage{nomencl}
\usepackage{etoolbox}
\usepackage[style=ieee, backend=biber, doi=false, maxnames=2, minnames=1, url=false]{biblatex}
\def \P {\boldsymbol{P}}
\def \Q {\boldsymbol{Q}}

\def \Y {\mathbf{Y}}
\def \V {\mathbf{V}}
\def \J {\mathbf{J}}

\def \R {\mathbb{R}}

\def \s {\boldsymbol{s}}
\def \sVec {\mathbf{s}}
\def \c {\mathbf{c}}

\def \Sigs {\mathbf{\Sigma}_{\s}}

\def \lamVec {\mathbf{\lambda}}
\def \setS {\mathcal{S}}
\def \setC {\mathcal{C}}
\def \wc   {\text{WC}}
\def \outsc {\text{OUT}}
\def \perfsc {\text{PERF}}

\newcommand{\RMSS}{Risk-Managed Steady-State Analysis}
\newcommand{\MCS}{Monte Carlo Simulation}

\newcommand{\nom}{\text{NOM}}
\newcommand{\op}{\text{OP}}

\newcommand{\crit}{critical performance measure}
\newcommand{\crits}{critical performance measures}
\newcommand{\Crits}{Critical performance measures}
\newcommand{\Stochp}{Stochastic parameters}
\newcommand{\stochp}{stochastic parameter}
\newcommand{\stochps}{stochastic parameters}

\newcommand{\Iline}{I_{\mathit{line}}}

\begin{document}

\title{A Risk-Managed Steady-State Analysis to Assess the Impact of Power Grid Uncertainties}

\author{
\IEEEauthorblockN{Naeem Turner-Bandele, ~\IEEEmembership{Member,~IEEE}, Amritanshu Pandey, ~\IEEEmembership{Member,~IEEE}, Larry Pileggi,~\IEEEmembership{Fellow,~IEEE}}

}



\maketitle

\begin{abstract}
\input{sections/abstract}
\end{abstract}

\begin{IEEEkeywords}
    \input{sections/keywords}
\end{IEEEkeywords}

\input{sections/nomenclature.tex}

\input{sections/introduction.tex}

\input{sections/background.tex}

\input{sections/rmss.tex}

\input{sections/results.tex}

\input{sections/conclusion}

\printbibliography


\end{document}

%% file: sections/abstract.tex
Electricity systems are experiencing increased effects of randomness and variability due to emerging stochastic assets. The increased effects introduce new uncertainties into power systems that can impact system operability and reliability. Existing steady-state methods for assessing system-level operability and reliability are primarily deterministic, therefore, ill-suited to capture randomness and variability. This work introduces a probabilistic steady-state analysis inspired by statistical worst-case circuit analysis to evaluate the risk of operational violations due to stochastic resources. Compared to parallelized Monte Carlo analyses (MCS), we have seen up to 24x improvement in runtime speed using our approach without significant loss of probabilistic accuracy for a Texas7k low-wind day test system.

%% file: sections/keywords.tex
probabilistic load flow, renewable generation, risk analysis, stochastic resources,  uncertainty analysis

%% file: sections/nomenclature.tex
\makenomenclature
\renewcommand\nomgroup[1]{%
  \item[\bfseries
  \ifstrequal{#1}{A}{Sets}{%
  \ifstrequal{#1}{B}{Variables and Parameters}{%
  \ifstrequal{#1}{C}{Abbreviations}{}}}%
]}

\nomenclature[A, 01]{\(\setC\)}{Set of \crits{}, indexed by $i=\{1,2,...,m\}$}
\nomenclature[A, 01]{\(\setS\)}{Set of \stochps{} (random variables), indexed by $i=\{1,2,...,n\}$}

\nomenclature[B, 01]{\(c_i\)}{$i^{th}$ \crit{}, $c_i \in \R$}
\nomenclature[B, 02]{\(\s_i\)}{$i^{th}$ \stochp, $\s_i \sim \left\{\mu_{\s_i}, \sigma_{\s_i}^2\right\}$}

\nomenclature[B, 03]{\(\c\)}{Vector of \crits{} $[c_i, i \in \setC]$}
\nomenclature[B, 04]{\(\sVec\)}{Vector of \stochps{} $[\s_i, i \in \setS]$}
\nomenclature[B, 05]{\(\rho\)}{Confidence level}
\nomenclature[B, 06]{\(\s^{l}, \s^{u}\)}{lower and upper confidence interval for \stochp}
\nomenclature[B, 07]{\(\mathbf{\Lambda}\)}{An $m \times n$ matrix of the sensitivities of $\c$ with respect to $\sVec$}
\nomenclature[B, 08]{\(\lamVec\)}{Vector of sensitivities for $c_i$ with respect to $\sVec$}
\nomenclature[B, 09]{\(\Sigs\)}{$n \times n $ \stochps{} covariance matrix}
\nomenclature[B, 10]{\(V\)}{Bus voltage magnitude, $V \in \setC$}
\nomenclature[B, 11]{\(\Iline\)}{Transmission line current flow, $\Iline \in \setC$}
\nomenclature[B, 12]{\(\overline{\c}, \underline{\c}\)}{\crit{} max and min limits}


\printnomenclature

%% file: sections/introduction.tex
\section{Introduction}

Electricity systems are undergoing structural transformation. The structural transformation is partly caused by the need to manage increasing amounts of variable and uncertain assets \cite{website:IEA-2021, website:EIA-2019, website:EIA-2020, website:Tribune-2022, website:EC-2021}. Despite rising uncertainty and variability, electricity system operators primarily use deterministic methods to perform system-level steady-state analysis of power grids. Failure to develop comprehensive and comprehensible steady-state methods for evaluating and mitigating risks due to stochastic resources could severely impact electricity system operations. Power system operators need new models, methods, and tools that incorporate uncertainty to address the increasing level of risk.

Presently, operators use power flow analysis as one way to assess a system's deterministic steady-state operability and reliability. In operations, engineers run multiple power flows on a limited set of scenarios using predetermined contingency criteria to evaluate the risk of a system's \crits{} (e.g., bus voltage magnitudes or line flows) and support decision-making \cite{Vaahedi-Practical-Power}. After conducting the analyses, operators manually set more restrictive limits if a grid is exposed to too much risk or less if the constraints defined by the analysis are too rigid \cite{Vaahedi-Practical-Power}. These practices work well  for grids comprised of fixed and predictable sources of uncertainty, but they are less effective for power grids that operate more stochastically.

\IEEEpubidadjcol

Only by using probabilistic methods can operators properly quantify risks from uncertainties. If operational practices fail to transition to probabilistic methods, the risk of system failures, blackouts, and near-misses could increase \cite{PNNL-StochasticOpPlanning}. Failure to transition could also result in increased real-time congestion problems, over-conservative operation limits, voltage instability, under-utilization of transmission assets, and excessive or insufficient operating reserves \cite{PNNL-StochasticOpPlanning, NERC-Integrating-Inverter}. The high cost of these risks requires novel modeling and analysis methods to quantify, evaluate, and mitigate the impact of resource variability to aid operators in their efforts to ensure operability, reliability, and resiliency \cite{USDOE-SG-2018}.

\begin{figure}[ht]
  {
    \begin{center}
      \includegraphics[trim=50 65 50 30,clip,width=0.75\columnwidth]{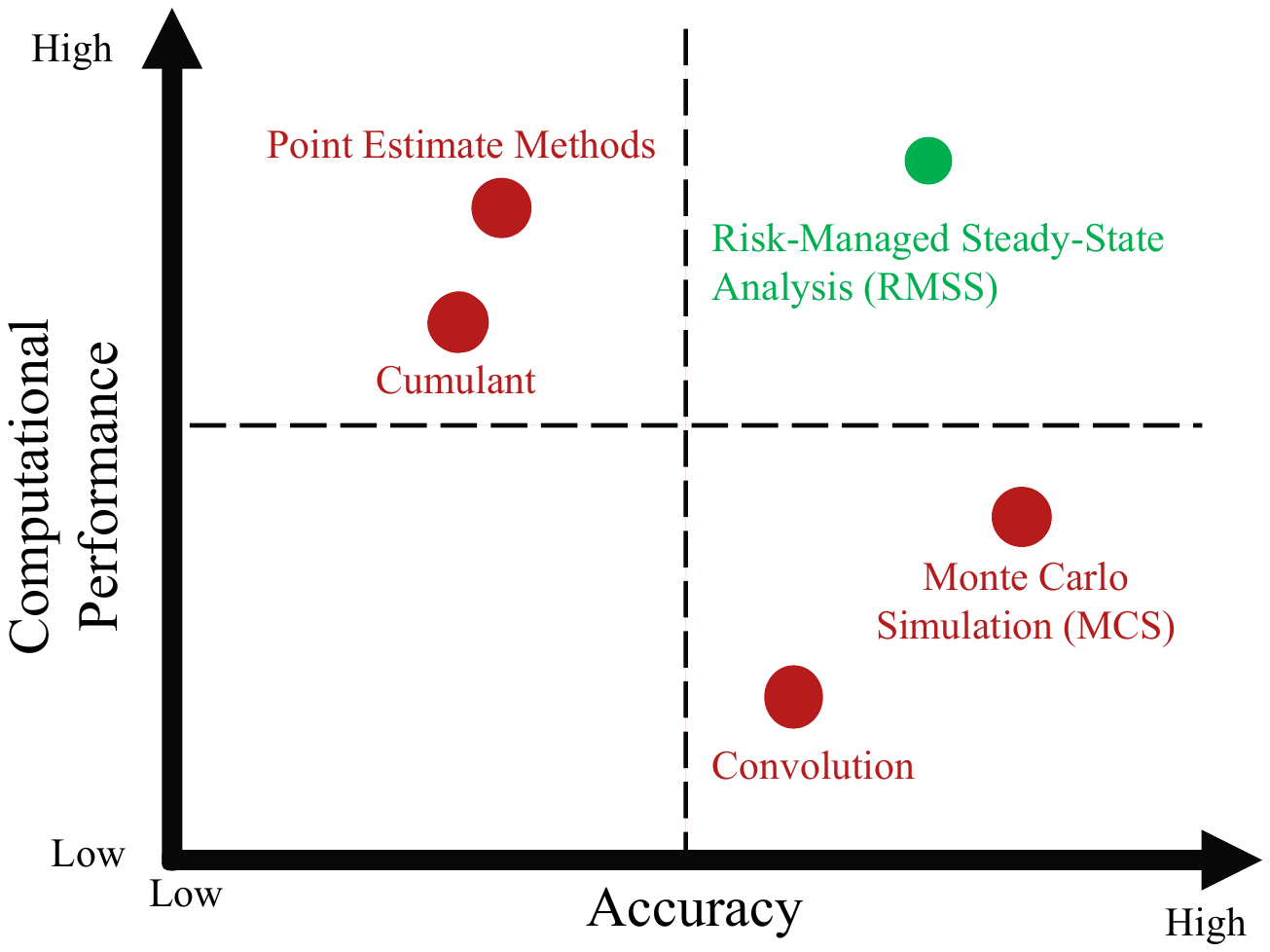}
    \end{center}
  }
    \caption{A comparison of probabilistic steady-state analyses in terms of accuracy and performance.}
    \label{fig:plf-comparison}
\end{figure}

Probabilistic load or power flows (PLF) are the most prevalent stochastic steady-state analyses in power systems. Probabilistic power flows consider the randomness and uncertainty associated with power system parameters. The objective is to represent all of the uncertainties present in a power system, model uncertainty propagation, and quantify the impact of uncertainties. Several techniques exist for performing PLFs. Broadly, the methods fall into three categories: numerical, analytical, and approximative. A four-quadrant plot qualitatively comparing the accuracy versus the performance of various methods is in Fig. \ref{fig:plf-comparison}. We compare and contrast the three methods in Section \ref{PLF}. 

Alternative approaches for steady-state probabilistic analyses exist in other fields (Section \ref{statistical-wc-analysis}). In integrated circuits, fluctuations in the manufacturing process can result in variations in the performance of an integrated circuit. Designers introduced worst-case analyses to better account for manufacturing fluctuations and to ensure a more stable circuit performance. A worst-case analysis seeks to identify the point(s) in a parameter space that might cause circuit performance to reach an extreme (corner) \cite{Orshansky-2008}.  Various extensions and iterations of worst-case circuit analysis have been used with great success for decades by engineers \cite{Nassif-1986, Director-1994, Orshansky-2008, Zhang-2009}, and producing integrated circuits would not be possible without the corresponding industry-standard tools that are used to capture manufacturing variations \cite{Fitzpatrick-2017}.

We introduce a probabilistic steady-state analysis to assess the risk of operational violations due to the presence of stochastic grid resources that we will refer to as a \RMSS{} (RMSS). In RMSS, we predict worst-case \crits{} $c$, compute the combination of \stochps{} $\s$ that produce predicted \crits{}, and process possible \crit{} violations. \Crits{} include bus voltage magnitudes and transmission line flows. The \stochps{} that we will consider are the active power of renewable generators and the active and reactive power of stochastic loads, but in general, our approach can accommodate the stochastic effects from any grid component model. Section \ref{rmss} describes our approach in detail.

We demonstrate the efficacy of our approach in terms of accuracy and performance by applying our method to a Texas7k low wind day test system in Section \ref{results}. We validate our method against \MCS{} (MCS) with 10,000 samples. Our results for the Texas7k low wind day test system show that:
\begin{enumerate}
    \item RMSS provides a good estimate of worst-case \crits{} when the mean critical performance is reasonably estimated
    \item RMSS runtime speedups of up to 24x when compared to parallelized MCS using up to 32 CPU cores
\end{enumerate}
We discuss the insights gained from our approach in Section \ref{discussion}.
\subsection{Contributions}
To summarize, our work provides the following contributions and results:
\begin{enumerate}
    \item A statistical steady-state analysis method that can assess the risk of operational violations due to stochastic resources (RMSS)
    \item Significant runtime speedup for a low wind day test case similar to those for which analysts need to make quick decisions
    \item An approximation with good accuracy that conservatively predicts the worst-case \crits{} when the mean \crit{} is adequately estimated
    \item Results that identify several possible \crits{} operational violations that could occur on low wind days in a synthetic network designed to mimic the ERCOT grid for which deterministic methods fail to capture the violations  
\end{enumerate}

The proposed RMSS can help power systems operators more adequately assess the increased operational risk to power systems due to emerging stochastic assets.

%% file: sections/background.tex
\section{Probabilistic Steady-State Analyses}
\label{Background}

\subsection{Traditional Probabilistic Load Flows} \label{PLF}

\input{sections/plf}

\subsection{Statistical Worst-Case Circuit Analysis} \label{statistical-wc-analysis}
\input{sections/statistical-wc-analysis}

%% file: sections/plf.tex
In the subsequent paragraphs, we summarize the standard numerical, analytical, and approximation PLF methods and discuss their efficacy in terms of accuracy and performance. We find that these methods vary widely in accuracy and performance, as shown in Fig \ref{fig:plf-comparison}. For comparison, we also show the efficacy of the proposed \RMSS{} (RMSS) discussed in Section \ref{rmss}.

Numerical PLFs draw repeated samples from probability density functions (PDF) of random variables and then perform deterministic power flows on each set of random samples. The most common numerical PLF method is Monte Carlo simulation (MCS) \cite{Tsai-2020, PNNL-StochasticOpPlanning}. Monte Carlo simulation is the most accurate PLF method and is widely considered the best in terms of accuracy. As such, MCS is the benchmark for other more recent approaches \cite{Prusty-2017, Zhang-2004}. 

Still, millions of simulations might be needed to achieve a high level of accuracy, and the computational performance of MCS suffers from a linear time complexity \cite{ uncertainty-morgan-henrion}.  The performance drawbacks make it challenging to perform Monte Carlo simulation on large networks or networks containing tens of thousands of random variables \cite{Ramadhani-2020, Prusty-2017, PNNL-StochasticOpPlanning}.  Monte Carlo simulation is impractical for operations since engineers may have to run N-1 PLF contingencies for millions of samples. Other methods offer better performance at the cost of lower accuracy.

Analytical methods use arithmetic operations on input PDFs to obtain PDFs for desired outputs. These methods simplify the power flow equations by linearizing them at an expected operating point \cite{ Ramadhani-2020, Prusty-2017} to achieve their results. The two most well-known methods are the convolution \cite{allan-daSilva-1981, Zhang-2004} and cumulant techniques \cite{Zhang-2004, Fan-2012}.

The convolution technique linearizes the power flow equations about the mean and then transforms inputs into desired outputs such as voltages and line flows using convolution \cite{allan-daSilva-1981, Zhang-2004}. The convolution method assumes independent random variables. 

Convolution methods suffer from several weaknesses.  Convolution methods are computationally intensive and require substantial storage and time to compute a solution \cite{ Zhang-2004, Fan-2012}. Also, they are inaccurate in the tail regions when inputs have large variances since the tail regions are far from the linearization point about the mean \cite{allan-daSilva-1981}. Alternative convolution approaches exist but use non-physical DC power flow equations \cite{Wang-Zhang-2017, Zhang-Kang-2016} and have yet to demonstrate their performance or accuracy for realistic systems.

\IEEEpubidadjcol

Conversely, the cumulant PLF method replaces the moments of a probability distribution with the cumulants. Cumulant methods use a linear relationship between input and output sensitivities combined with arithmetic operations to compute the cumulants and moments of the output probability distributions \cite{Zhang-2004, Fan-2012}. Cumulant PLF methods use expansions to obtain PDFs of output RVs. Cumulant methods perform better than convolution methods. Disadvantages of cumulant methods are that they are less accurate when correlations are considered \cite{Prusty-2017} and accuracy suffers when the order of the cumulants is truncated \cite{Fan-2012, Aien-2016}.

Approximation techniques use the statistics of the inputs of a probability distribution to assess critical locations of interest to execute deterministic power flows and then estimate the output distribution variables’ statistics. The most well-known techniques are point estimate methods (PEM) which use the first several central moments of input random variable distributions to identify sample concentrations for all random variables. Concentrations are pairs that consist of a (1) random variable sample location where a deterministic power flow is run and (2) a weight to measure the impact of the power flow evaluation at the location on the behavior of output variables \cite{Su-2005, Morales-2007, Morales-2010, Che-2020}.

Approximation methods are considered quick, but they cannot develop accurate estimates for higher-order statistics beyond the mean, are less accurate as the coefficient of variation increases, and are less successful when the input distribution is highly asymmetrical \cite{Ramadhani-2020, Prusty-2017, Zhou-2019, Che-2020}. Approximation methods are some of the most used in literature \cite{Che-2020, Wu-2015, Mohammadi-2013, Morales-2010, Morales-2007, Su-2005}; however, almost none of the works in the literature apply approximation methods to realistic networks, so their efficacy is unclear.

%% file: sections/statistical-wc-analysis.tex
Tangential to power grids, methods exist in the circuit simulation domain to perform probabilistic steady-state analyses. An approximation method in the circuit simulation domain is statistical worst-case analysis. A statistical worst-case analysis predicts the limits of circuit performance and identifies the components in a circuit that are critical to circuit performance \cite{Director-1994, Dharchoudhury-1995, Sengupta-2005, Orshansky-2008}. 

\begin{figure}[htbp]
  {
    \begin{center}
      \includegraphics[trim=10 65 10 50,clip,width=0.75\columnwidth]{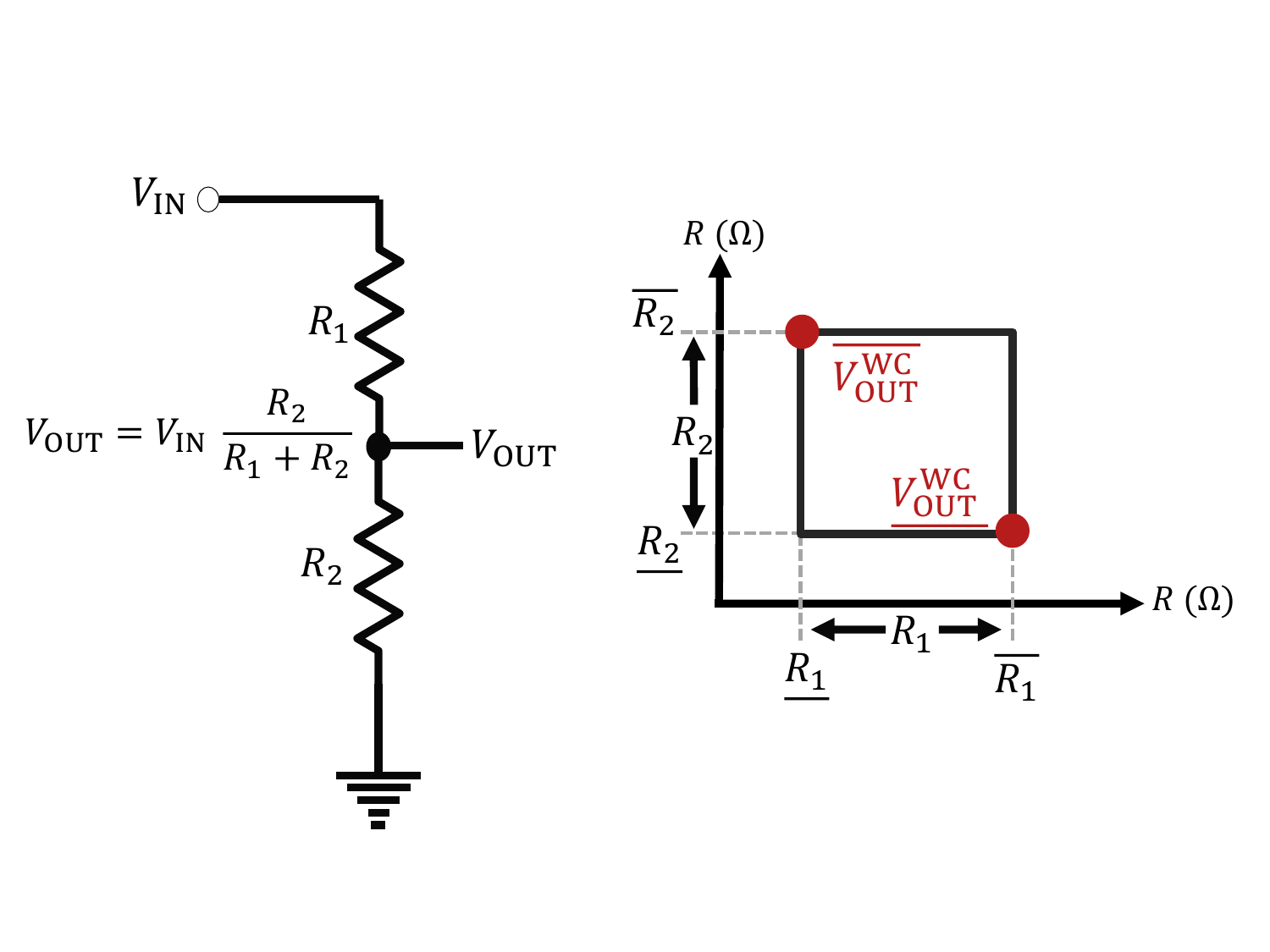}
    \end{center}
  }
    \caption{An example of worst-case analysis using a voltage divider. $V_{\outsc}$ hits an upper bound $\overline{V^{\wc}_{\outsc}}$ when $R_{2}$ is at its maximum resistance and $R_{1}$ is at its minimum resistance. $V_{\outsc}$ reaches a lower bound $\underline{V^{\wc}_{\outsc}}$ when the opposite is true.}
    \label{fig:wc-example}
\end{figure}

To illustrate the concept of worst-case analysis, consider the resistive voltage divider circuit in Fig. \ref{fig:wc-example} \cite{Orshansky-2008}. We assume that our parameters, the discrete resistors $R_{1}$ and $R_{2}$, have a tolerance associated with them that represents an upper and lower bound for the resistance (e.g., $\pm 5\%$). We can model these resistors as random variables that result in the parameter space in Fig. \ref{fig:wc-example} \cite{Orshansky-2008} when placed together. For this simple circuit, one possible performance measure is the output voltage, $V_{\outsc}$. 

A worst-case analysis would seek to identify the resistance values that result in $V_{\outsc}$ reaching an extreme. There are two limits for $V_{\outsc}$. $V_{\outsc}$ is largest (denoted by $ \overline{V^{\wc}_{\outsc}} $) when $R_{2}$ is at its maximum resistance ($\overline{R_2}$) and $R_{1}$ is at its minimum (denoted by $\underline{R_1}$). $V_{\outsc}$ is smallest ($\underline{V^{\wc}_{\outsc}}$) when $R_{1}$ is at its maximum ($\overline{R_1}$) resistance and $R_{2}$ is at its minimum ($\underline{R_2}$). While we can obtain trivially worst-case metrics for this simple circuit, we now discuss how worst-case metrics can be obtained for generic circuits.

\subsubsection{Applying Worst-Case Analysis to Integrated Circuits}

To apply worst-case analysis to integrated circuits, designers redefined the worst performance as the performance that bounds some desired percentage of performances (e.g., the 75th percentile) \cite{Director-1994}. With that revised definition in mind, the objective now becomes to identify the \stochps{}, $\s$, that guarantee that the likelihood of the circuit's \crit{}, $c$, is no worse than the worst-case \crit{}, $c^{\wc}$, is probability $\rho$. 

A worst-case analysis consists of two phases, construction and analysis \cite{Director-1994}. The construction phase identifies the worst-case \stochps, and the analysis phase simultaneously evaluates the circuit at those worst-case \stochps. In the subsequent subsections, we will explain both steps.

\subsubsection{Construction Phase}
We begin with a circuit whose \stochps, $\s$, make up a performance function $\c$. Let $\c$ be an $m \times 1$ vector of critical performances that we want to evaluate, and let $\sVec$ be an $n \times 1$ random vector of \stochps. The mathematical definition of such a performance function is:
\begin{equation}
    \c = f_{\perfsc}(\sVec)
    \label{performance-measure}
\end{equation}
In integrated circuits, the idea is that in a well-controlled manufacturing process, a circuit's \stochp{} variations are on the order of a few percent, so one can approximate a performance equation using a truncated first-order Taylor Expansion. The idea that the performance can be well-approximated by a truncated first-order Taylor Expansion highlights the first of two fundamental assumptions of any worst-case analysis. This assumption is that monotonicity exists between our device \stochps{} and circuit \crits. If monotonicity does not hold, we would need to perform multiple evaluations of our circuit.

Assuming that monotonicity holds, we take a first-order Taylor expansion of  \eqref{performance-measure} and obtain:
\begin{equation}
    \c \approx \c^{\nom} + \mathbf{\Lambda} (\sVec - \sVec^{\nom})
\end{equation}
where $\s^{\nom}$ is our nominal set of \stochps, $\mathbf{\Lambda}$ is a $m \times n$ matrix of the sensitivities of our \crits{} with respect to our \stochps, and $\c^{\nom} = f_{\perfsc}(\sVec^{\nom})$. 
In worst-case analysis, we look at one \crit{} at a time. So, rewriting the above for the $i^{th}$ \crit{}, we obtain:
\begin{equation}
\label{eq:wc_hyp}
    c_{i} = c^{\nom}_{i} + \lamVec(\sVec -\sVec^{\nom})
\end{equation}
Here, $c_{i}$ is a scalar quantity and our $i^{th}$ \crit{}, and $\lamVec$ is a vector consisting of the sensitivities of our \crit{} with respect to our \stochp. 

The second fundamental assumption of a worst-case analysis is that we assume our \stochps{} $\sVec$ can be characterized by or approximated using a multivariate normal distribution with a vector of $n \times 1$ means $\sVec^{\nom}$ and an $n \times n $ covariance matrix $\Sigs$.

\[ \sVec \in N(\sVec^{\nom}, \Sigs)
\]

We can estimate the upper and lower bounds of $c^{\wc}_{i}$ by calculating the following \cite{Director-1994, Koski-2017}:
\begin{equation}
\label{eq:wcperf}
	c_{i}^{\wc} = c^{\nom}_{i} \pm \Phi^{-1}(\rho)\sigma_{c_{i}}
\end{equation}
where $c^{\nom}_{i}$ is the nominal or mean performance value, $\Phi^{-1}$ is the inverse standard normal CDF $(\mu=0, \sigma=1)$, $\sigma_{c_i} = \sqrt{\lamVec^T \Sigs \lamVec}$ is the standard deviation of the performance, and $\pm$ is the worst-case direction. \cite{cacuci-2003, cacuci-2018} provides a general proof for obtaining $\sigma_{c_i}^2$ using the propagation of moments methods.

If we set  value  for $c_{i}^{\wc}$ equal to \eqref{eq:wc_hyp}, then \eqref{eq:wc_hyp}, is a hyperplane. Any combination of \stochps{} along the hyperplane can produce $c_{i}^{\wc}$. To find the true worst-case, we restrict our search space by looking for the most probable value in the hyperplane. We can formulate this problem as a linear program \cite{Director-1994}:
\begin{equation}
	\label{eq:wcopt}
	\begin{aligned}
    & \underset{\s}{\text{minimize}}
    & & (\sVec - \sVec^{\nom})^T \Sigs^{-1} (\sVec - \sVec^{\nom}) \\
    & \text{subject to}
    & & \lamVec(\sVec - \sVec^{\nom}) + c^{\nom}_{i} = c^{\wc}_{i}
	\end{aligned}
\end{equation}
The convex optimization problem in \eqref{eq:wcopt} tells us that we seek the most probable point on the hyperplane by examining the \stochps{} closest to the nominal in a multivariate normal distribution. The term $(\sVec - \sVec^{\nom})^T \Sigs^{-1} (\sVec - \sVec^{\nom})$ is effectively a normalized distance, telling us how close or far away from our nominal parameter we are. If the parameters are well correlated, dividing by the covariance results in a small distance. Likewise, the opposite is true as well. 

After deriving the Lagrangian function, obtaining the KKT conditions, solving for our primal and dual variables, and then using algebraic manipulation, we obtain the closed-form solution for the worst-case \stochps:
\begin{equation}
	\label{eq:wcparam}
	\sVec^{\wc} = \sVec^{\nom} + \frac{c_{i}^{\wc} - c^{\nom}_{i}}{\lamVec^{T} \Sigs \lamVec}\Sigs \lamVec
\end{equation}
Since our optimization problem is convex, any local optimal point we obtain is also globally optimal.

\subsubsection{Analysis Phase}
The analysis phase is trivial. Upon obtaining our worst-case circuit parameters, $\sVec^{\wc}$, we perform a simulation of our circuit simultaneously at all worst-case \stochps. The analysis phase concludes by ensuring that the simulated performances at the worst-case \stochps{} match our desired confidence interval.

%% file: sections/rmss.tex
\section{\RMSS{} (RMSS)} \label{rmss}
We next describe our approach for \RMSS. We begin with an example that illustrates the need for RMSS, then redefine the statistical worst-case analysis problem from Section \ref{statistical-wc-analysis} into a risk-management problem.

\input{sections/rmss/rmss-need.tex}

\input{sections/rmss/rmss-problem.tex}

\input{sections/rmss/monotonic-assumption.tex}

\input{sections/rmss/normal-assumption.tex}

\input{sections/rmss/sensitivity-analysis.tex}

\input{sections/rmss/implementation.tex}

%% file: sections/rmss/rmss-need.tex
\subsection{The Need for \RMSS} 
\begin{figure}[h!t]
  {
    \begin{center}
      \includegraphics[trim=100 20 100 20,clip,width=0.6\columnwidth]{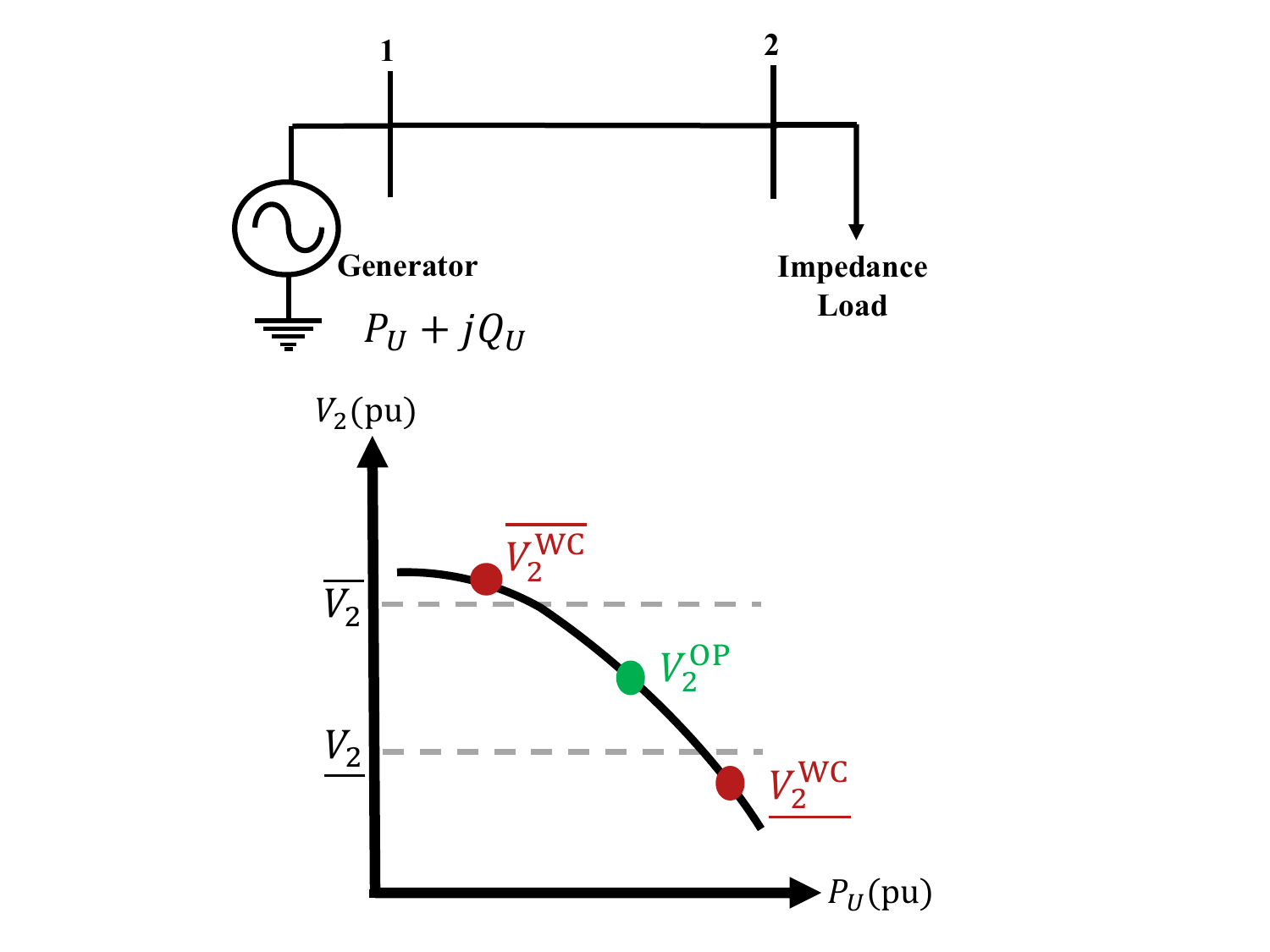}
    \end{center}
  }
    \caption{An example of a worst-case analysis for a two-bus power system with a single stochastic generator and a fixed impedance load.}
    \label{fig:wc-power-systems-example}
\end{figure}

To illustrate the need for RMSS, we examine the two-bus power system with a single stochastic generator and an impedance load in Fig. \ref{fig:wc-power-systems-example}. Our impedance load in this example is a fixed known quantity. 

Imagine that an operator of this two-bus grid plans to dispatch the stochastic generator a day ahead to a mean value $P_U$ based on forecasts. The stochastic generator is intermittent and variable, and no forecast is perfect. Some day-ahead (DA) forecast error is likely due to forecast uncertainty. Our operator is unsure that the stochastic generator will achieve the expected value $P_U$. In this scenario, the operator wants to answer one fundamental question: Will the uncertainty of $P_U$ result in the voltage at bus 2 $V_2$ violating its limits?

The operator could adopt two approaches to solve the problem and answer the critical question. The first option is to perform a deterministic analysis for which the operator runs a power flow with $P_U$ at its forecasted value while all other system parameters and variables remain at their nominal values. The second option is to perform a worst-case analysis. We next consider and compare results from both approaches below.

If the operator performs a deterministic analysis, they will find that there is an operating point, $V^{\op}_{2}$ that is within desired voltage limits $\overline{V_{2}}$ and $\underline{V_{2}}$. Here, we assume that the grid operator determines $\overline{V_{2}}$ and $\underline{V_{2}}$ based on engineering judgment and system knowledge. Alternatively, an ISO, utility, or other organization might know these limits based on the transient analysis study. The deterministic analysis would conclude by informing the operator that all is well and that there is no risk of exceeding the voltage bounds.

If the operator performs a worst-case analysis, they would see the limits of deterministic analysis. In a worst-case analysis, the operator models the stochastic generator's active power, $\P$, as a normally distributed random variable $\mathcal{N}(\mu_P, \sigma_P^2)$. The worst-case analysis might show that due to DA forecast errors, a particular dispatch of $P_U$ (drawn from the distribution of $\P$) can result in a worst-case upper bound $\overline{V^{\wc}_{2}}$ or a worst-case lower bound $\underline{V^{\wc}_{2}}$ that violates the system voltage limits. This might indicate that the operator should take corrective action to avoid the possibility of voltage violations. 

Deterministic analysis for an $N$-bus system with $M$ random variables would be more challenging and time-consuming. In an $N$-bus system with $M$ stochastic generators, it would be difficult to determine the worst-case bounds of \crits{} such as bus voltage magnitudes and line flows or estimate the operational violation risk to the power grid. Moreover, if an operator needs to also perform a contingency analysis for the $N$-bus system, they will require extensive computational resources to solve the problem quickly.

A worst-case analysis would help a grid analyst mitigate the risk of operational violations. By performing a worst-case analysis, a grid analyst can see the range of \crits, determine if they violate operating limits, and know the \stochps{} contributing to an operational violation. Identifying the location of the operational violation limits and \stochps{} can help grid analysts target mitigation measures. Equipped with this knowledge, a grid analyst can more effectively deploy operational violation mitigation measures such as aggregating renewable plants, deploying energy storage, employing load control, or securing additional reserves.

An RMSS predicts the worst-case \crits, estimates the combination of \stochps{} that produce predicted \crits{}, and identifies possible violations, which might remain unobserved during deterministic analysis. Decision-makers can use RMSS to make targeted and effective decisions to mitigate operational violations. With the value of RMSS analysis clear, we proceed to extend the worst-case analysis problem and define the risk-managed steady-state analysis problem.

%% file: sections/rmss/rmss-problem.tex
\subsection{The Risk-Managed Steady-State Analysis Problem} 

Consider an electric power grid model consisting of buses connected by branches and transformers with a mix of stochastic and deterministic generators and loads. A deterministic generator model represents those that are fueled by coal, oil, natural gas, or nuclear energy. We can dispatch deterministic generators readily when called. In contrast, a stochastic generator represents variable and uncertain energy sources such as wind or solar. Loads operate similarly. A deterministic load might be a commercial business or industrial plant, while a stochastic load might be an aggregation of electric vehicle charging stations in a future grid scenario. 

We can forecast the power we expect stochastic generators to supply or stochastic loads to demand. System operators use forecasts to predict generation and load days in advance and up to 5 minutes in advance. The 5-minute forecasts for stochastic generation and load are reasonably accurate, but day-ahead forecasts are imprecise at best \cite{Mauch-2013, Hodge2013, Apt-2014}. Aggregation of forecast errors can mask the imprecision of day-ahead forecast errors at individual levels \cite{HABEN20191469, Jacob-2020, Souhaib-2021}. 

\begin{figure}[h!t]
  {
    \begin{center}
      \includegraphics[width=0.7\columnwidth]{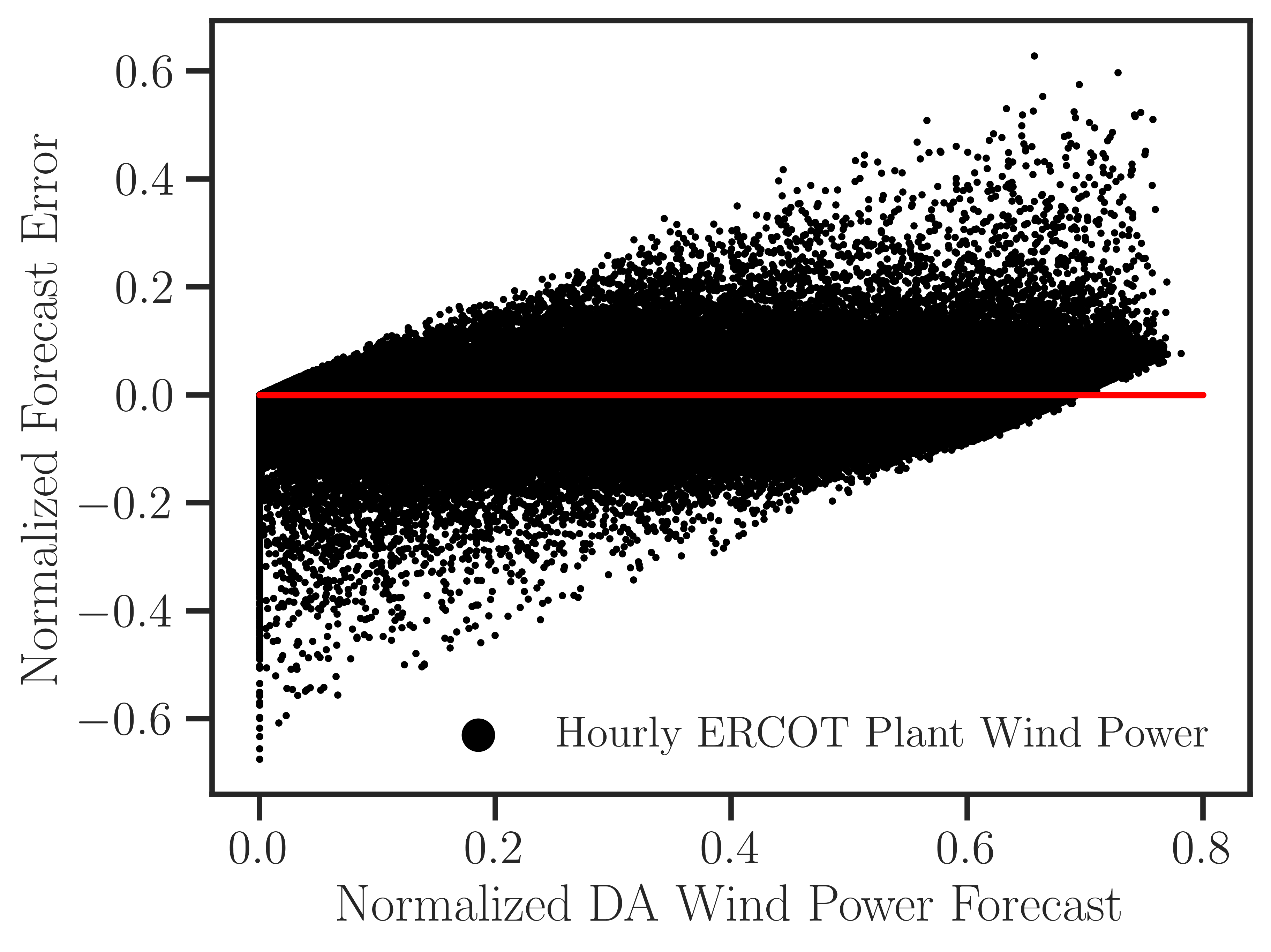}
    \end{center}
  }
    \caption{The normalized day-ahead (DA) ERCOT wind power forecast errors at 125 existing wind sites plotted against the wind power forecasts from June to August 2018.}
    
    \label{fig:wind-forecasts-errors}
\end{figure}

To demonstrate the inaccuracy of day-ahead forecasts, we plot the normalized day-ahead ERCOT wind power forecast errors against the wind power forecasts from June to August 2018 in Fig. \ref{fig:wind-forecasts-errors}. The data presented uses 125 existing wind sites. The National Renewable Energy Laboratory (NREL) created the forecast and actual synthetic wind power data. \cite{PERFORM-Forecasts} describes the methods used to create the data. 

The wind forecast error ($e$) is the forecast value ($F$) minus the actual wind power ($A$), stated as $e = F - A$. The installed wind capacity normalizes the forecast and actual wind power. Negative forecast errors represent under-forecasts, while positive errors represent over-forecasts. The plots show that forecasts systematically under-predict or over-predict the amount of wind power. The inaccuracy of forecasts illustrates the need for probabilistic analyses.

DA forecasts for stochastic generators and loads will never be exact. Stochastic generators and loads are crucial to system operation today and will continue to be crucial moving forward. If stochastic generation and demand deviate significantly from forecasts, then the risk of operational violations increases. RMSS aims to help grid analysts quantify the risk of \crit{} operational violations due to forecast errors and identify the \stochps{} that produce operational violations.

We solve the operational violation problem by determining the risk of \crit{} violations and identifying the worst-case \stochps. We identify the worst-case \stochps{} by predicting how DA forecast errors might affect select \crits. \Stochp{} could be the $P$ of renewable generators and the $P$ and $Q$ of loads. \Crits{} would include bus voltage magnitudes and line flows, but additional performance measures are possible. A \crit{} violation is a \crit{} outside its limits. For example, line flows being above their maximum limits or bus voltage magnitudes being above or below their maximum and minimum limits. We find worst-case \crits{} using \eqref{eq:wcperf} and worst-case \stochps{} using \eqref{eq:wcparam}.

Given the variability associated with stochastic generators and loads, we can model their real and reactive powers as real continuous random variables $\P$ $\sim \left\{\mu_{\P}, \sigma_{\P}^2\right\}$ and $\Q$ $ \sim \left\{\mu_{\Q}, \sigma_{\Q}^2\right\}$. Here, $\mu$ is the mean, and $\sigma$ is the standard deviation. Depending on the stochastic generation or load type, the RV has a specific distribution due to its DA forecast error.

RMSS uses \eqref{eq:wcperf} and \eqref{eq:wcparam} to obtain the worst-case \crits{} $c$ and \stochps{} $\s$. To use these two equations, we must check the monotonicity and normality assumptions apply to the steady-state analysis of power systems.

%% file: sections/rmss/monotonic-assumption.tex
\subsection{Monotonicity Assumption}

We consider the monotonicity assumption first.  In a monotonic relationship, variables move in the same relative direction but not necessarily at the same rate meaning monotonic functions can be linear or nonlinear \cite{TheConciseOxfordDictionaryofMathematics}. A monotonic relationship between \stochps{} and \crits{} is crucial since it guarantees that extremes exist for our \crits. Without extreme limits, it is impossible to obtain worst-case \crits.

For our steady-state analysis (power flow) of RMSS, we utilize the equivalent circuit formulation (ECF) from \cite{SUGAR-PF}. The power flow equations in this formulation are built on KCL-based nodal equations and are nonlinear. 

We can represent the power grid abstractly via a linear circuit at its operating point in ECF. In our linear circuit, loads and generators are modeled by equivalent impedances, and the slack node is modeled as an independent voltage source. All elements in this circuit are fixed except for the nodal injection from \stochps{} that we aim to change. 

In the remainder of this section, we assume a linear circuit representation of the power grid in a stable operating region as we discuss if the change in \crits{} with respect to \stochps{} is monotonic. We also assume that there are reasonable bounds on how much \stochps{} can change.

The first relationship we evaluate for monotonicity is the one between \crit{} node voltage $V$ and our \stochps{} real and reactive power injections $P$ and $Q$.  If there is no negative resistance, it is trivial to show that any increase in real power injection will result in a larger voltage magnitude drop across the series and lower nodal voltages for our linearized circuit. 

Reactive power injections in the stable operating region \cite{Kundur1994} are multifaceted and depend on whether we inject inductive or capacitive reactive power. If we make the source impedance more capacitive (inject more leading reactive power), we will observe an increase in nodal voltages throughout the circuit. On the other hand, if we make the source more inductive (inject more lagging reactive power), we will observe a decrease in nodal voltages. 

In general, we find that existing works in power systems analysis also support the node voltage magnitude monotonic assumption under moderate changes in \stochps{}. P-V and Q-V curves in the standard operating range are generally monotonic \cite{Kundur1994}.

The second relationship we assess is the one between \crit{} line flow current magnitude $I_{line}$ and our \stochps{}, the real and reactive power injections $P$ and $Q$. In RMSS, we are only concerned with increases in line flow current magnitudes that result in a line flow exceeding its rated limit. 

The flow of real power between two buses depends on the line impedance, sending-end voltage magnitude, receiving-end voltage magnitude, and the angle difference between sending and receiving bus voltages \cite{Kirtley-2010}. Changing voltage magnitudes (within operating range) has little impact on the real power flow across the line. The sign of the difference in voltage phase angle between sending and receiving nodes is the only variable that, if flipped, could produce a nonmonotonic relationship between line flow current magnitudes and real power injections. So long as the angular difference does not change in sign, real power injections are monotonic with respect to line flow current magnitudes.

Relatedly, the flow of reactive power on a line depends on the line impedance, sending-end bus voltage magnitude, and the sending and receiving bus voltage phase angle \cite{Kirtley-2010}. Reactive power flow is most dependent on the difference between sending and receiving bus voltage magnitudes. If the change in stochastic parameters does not cause a change in the sign of voltage magnitude difference between sending and receiving nodes, then reactive power injections are monotonic with respect to line flow current magnitudes.  

Thus, we posit that the relationship between \stochps{} and \crits{} appears to be predominately monotonic within the operating range for moderate changes to stochastic parameters. Consequently, the monotonic assumption is reasonable, and it is possible to find worst-case \crits. Still, suppose the magnitude of change in current or power injected is large. In that case, the relationship between \stochps{} and \crits{} may not be monotonic, and any user of RMSS will need to validate the assumption. We further validate this assumption by comparing the RMSS approach against Monte Carlo simulations in Section \ref{Validation}.

%% file: sections/rmss/normal-assumption.tex
\subsection{Normality Assumption}
We examine the normality assumption since the monotonicity assumption is reasonable. The normality assumption states that we can model our \stochps{} as normal. Prior work \cite{Zhang-2015, Hodge-2011, Wang-2022, Hodge2013, Apt-2014, Mauch-2013} and our analysis shows that the normal distribution is not an exact fit.

The forecast error determines the distribution of a \stochp. The consensus is that wind, solar, and load forecast errors are nonnormal \cite{Zhang-2015, Hodge-2011, Wang-2022, Hodge2013, Apt-2014, Mauch-2013}. There is no consensus on the exact distribution of wind, solar, and load forecast errors. Our analysis agrees that forecast errors are generally nonnormal. 

\begin{figure}[h!t]
  {
    \begin{center}
      \includegraphics[width=0.7\columnwidth]{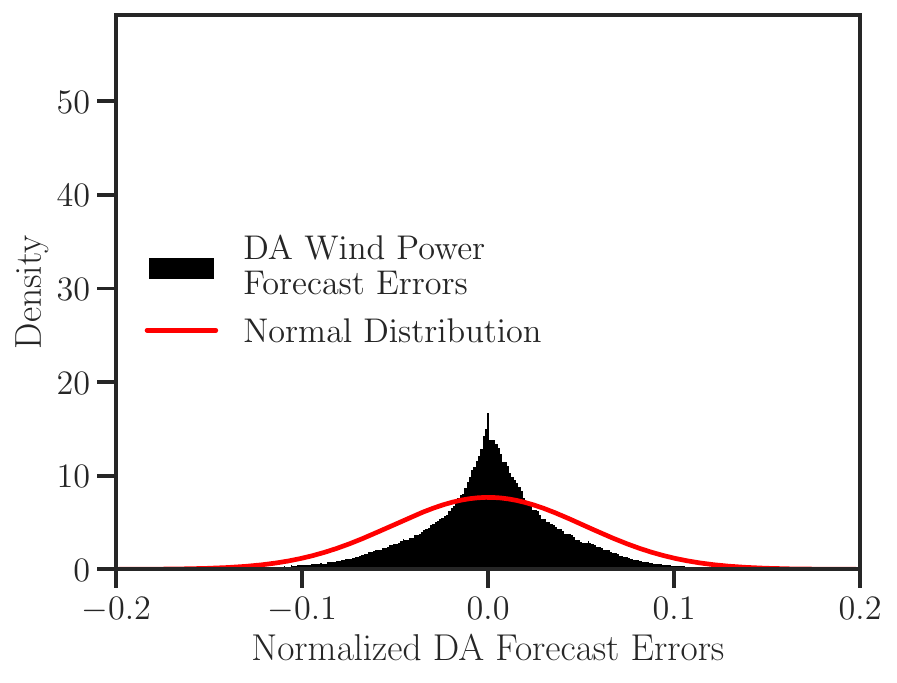}
    \end{center}
  }
    \caption{A histogram of the DA synthetic wind power forecast errors at 125 wind sites from June to August 2018.}
    \label{fig:nonnormal-wind}
\end{figure}

We analyzed the distribution of the DA synthetic wind power forecast error at 125 existing wind sites in ERCOT from June to August 2018. NREL created the forecast and actual synthetic wind power data. \cite{PERFORM-Forecasts} describes the methods used to create the data. A histogram of the June to August 2018 DA wind power forecast errors in ERCOT in Figure \ref{fig:nonnormal-wind} shows that the DA forecast error of wind is nonnormal. A normal distribution fails to capture the peak of the DA forecast error in ERCOT. 

The actual distribution of DA wind power forecast error is leptokurtic and left-skewed. A prior analysis of wind farms across ERCOT yielded similar results \cite{Mauch-2013, Apt-2014}. Examining data for DA load and solar forecast errors also shows that they are nonnormal \cite{Wang-2022, Apt-2014, Souhaib-2021}.

The distribution that best fits DA wind, solar, or load forecast errors is unknown. Some evidence has shown that a logit-normal \cite{Mauch-2013, Apt-2014} or hyperbolic \cite{Hodge2013} distribution are better fits for DA forecast errors.

Although we presume normality, the worst-case analysis should still provide meaningful results. The normal distribution will still capture a significant portion of a \stochp’s actual distribution. Further, a worst-case analysis assuming normality can provide more significant insights than a deterministic analysis and will result in a conservative worst-case estimate as compared with using the actual nonnormal distribution. Future work will explore the possibility of including a logit-normal distribution into our formulation or using transformations to restore normality.
 
%% file: sections/rmss/sensitivity-analysis.tex
\subsection{Sensitivity Analysis}
\label{sensitivity-analysis}

First-order sensitivities provide the maximum contributions to a \crit's variance/covariance, skewness, and kurtosis \cite{cacuci-2018}. Without sensitivities, performing any form of predictive modeling like RMSS is impossible\cite{cacuci-2018}. Thus, it is vital to obtain the first-order \crit{} sensitivities accurately and efficiently. 

Methods to obtain sensitivities can be either deterministic or statistical (global) \cite{saltelli-2007, cacuci-2018}. Statistical methods are easy to implement but may require thousands of simulations to obtain the exact sensitivities \cite{cacuci-2018}. Deterministic methods can rapidly get thousands of performance sensitivities but may suffer when the performance behavior is highly nonlinear. This work uses deterministic methods for computing sensitivities to maximize computational efficiency. 

We obtain sensitivities deterministically using adjoint sensitivity analysis (ASA). This method uses the adjoint system to compute the first-order sensitivities. Adjoint sensitivity analysis is used in circuits, nuclear engineering, and various other fields to estimate the first-order sensitivities quickly and accurately \cite{bakr-2013, Pileggi-1995, cacuci-2018}.

Our adjoint sensitivity analysis is summarized as follows. We begin by performing an ECF power flow analysis on our power grid which is the solution to the system of linear equations of the form:
\begin{equation}
\label{eq:solve-normal-system}
	\Y(x)\V = \J
\end{equation}
where $\Y$ is the $m \times n$ system matrix with parameters $x$, with $\s \in x$. $\V$ is a vector that contains our state variables and \crits{}, and $\J$ is the excitation vector. Then, we solve for the adjoint system to produce the sensitivities.

With $\frac{\delta \V}{\delta \J}$, the solution to our adjoint system, we can find the $c_i$ sensitivity with respect to the $i^{th}$ parameter $x_i$  using  \eqref{eq:compute-sensitivities} \cite{Pileggi-1995}. 
\begin{equation}
\label{eq:compute-sensitivities}
	\frac{\partial \V_i}{\partial x_i} = \left( \sum_{k,l} \frac{\partial \V_{i}}{\partial \Y_{k,l}}\frac{\partial \Y_{k,l}}{\partial x_i} + \sum_{k} \frac{\partial \V_{i}}{\partial \J_{k}} \frac{\partial \J_{k}}{\partial x_{i}}   \right)
\end{equation}
$k$ corresponds to the row of $\Y$ and $\J$. $l$ is the column of $\Y$. We continue solving until we obtain all our desired first-order sensitivities. A more detailed discussion of adjoint sensitivity analysis is in \cite{bakr-2013, Pileggi-1995}.

%% file: sections/rmss/implementation.tex
\subsection{Implementation}
 Any RMSS implementation starts by parsing inputs such as grid models, confidence intervals ($\rho$), \stochps, and \crits. Then, a power flow is run at the given initial conditions to estimate mean \crits. We use a single power flow evaluation to estimate mean \crits{} given that we lack knowledge of the sample or population means of \crits{}. After estimating mean \crits{}, we perform ASA to get all \crit{} sensitivities. If there are several thousand \crit{} sensitivities to compute, then we use parallel processing to get the sensitivities; otherwise, we compute sensitivities serially. Once we have the sensitivities, we predict all worst-case \crits{} using \eqref{eq:wcperf}. The mean \crit{} from power flow, $\rho$, and the approximate standard deviation $\sigma_{c_i} = \sqrt{\lamVec^T \Sigs \lamVec}$ are used to find $c_i^{\wc}$ $ \forall c_i \in \setC$. The penultimate step in RMSS is to estimate the $\sVec^{\wc}$ corresponding to a given $c_i^{\wc}$. Each $c_i^{\wc}$ has its own $\sVec^{\wc}$. RMSS concludes by processing all \crit{} violations.

We developed a Python and Cython implementation of this RMSS procedure. For demonstration purposes, we considered \crits{} for bus voltage magnitudes and line flows. \Stochp{} were incorporated for the $P$ of stochastic generators and loads and the $Q$ of stochastic loads.

\input{sections/rmss/rmss-algorithm.tex}

%% file: sections/rmss/rmss-algorithm.tex
\RestyleAlgo{ruled}
\SetKwComment{Comment}{/* }{ */}
\SetKw{Continue}{continue}
\SetKw{remove}{remove}
\SetKwInput{Read}{Read}
\SetKwInput{Run}{Run}
\SetKwInput{Create}{Create}
\SetKwInput{Initialize}{Initialize}
\SetKwInput{Solve}{Solve}
\SetKwInput{Estimate}{Estimate}
\SetKwInput{Compute}{Compute}
\SetKwInput{Predict}{Predict}
\SetKwInput{Process}{Process}
\SetKwInput{Store}{Store}

\begin{algorithm}[ht]
\caption{\RMSS{}}\label{alg:rm-ss}
\KwIn{Grid network models, initial conditions, $\s$, $(\overline{\c}=\{\overline{V}, \overline{\Iline}\}$, $\underline{\c}=\{\underline{V}\})$, and $\rho$}
\Solve{power flow at initial conditions to estimate $\c^{\nom}$}
\Run{ASA to obtain all $\lambda \in \Lambda$}

\Predict{all $\c^{\wc}$ using \eqref{eq:wcperf}}
\For{each $c_{i}^{\wc}$}{
    \Compute{$\sVec^{\wc}$ for $c_{i}^{\wc}$ using \eqref{eq:wcparam}}
}
\Process{all $\c$ violations $(c^{\wc} > \overline{c}, c^{\wc} < \underline{c})$ whose $\s \in (\s^{l}, \s^{u})$}
\end{algorithm}

%% file: sections/results.tex
\section{Results}\label{results}

We constructed several synthetic Texas7k summer 2018 low wind day test systems to compare the accuracy and computational performance of RMSS with MCS. The low wind day test systems are available at \cite{Texas7k-LW}. We selected low wind day test systems to assess recent concerns about low wind days and their impact on grid performance \cite{website:Tribune-2022}. In ERCOT, a low wind generation day in the summer is a day where summer wind generation is around 11\% of its capacity \cite{website:Tribune-2022}. A selection of the results is presented here for brevity's sake. Fig. \ref{fig:vmag-results} and \ref{fig:line-flow-results} compare the accuracy of RMSS to MCS. Fig. \ref{fig:perf-results} compares the computational performance of RMSS to MCS.



\subsection{Measuring Accuracy and Performance} \label{Validation}

Accuracy and computational performance are the metrics we used to evaluate the efficacy of the RMSS algorithm. 

The \crits{} analyzed were bus voltage magnitudes $V$ and the current magnitude of line flows $\Iline$. We assessed the accuracy of our work by obtaining the MCS 95\% confidence interval of the mean for all \crits{} and comparing the values obtained to the RMSS 95\% confidence interval \crit{} values. We plotted both sets of confidence intervals on scatter plots to ascertain their relationship. We hypothesized that a strong linear relationship between the two data sets would indicate that the RMSS solution mimics the MCS solution in the tail regions. We chose this approach instead of presenting an error metric to ensure we could see any outliers in the data.

We measured computational performance by calculating the speed improvement of RMSS over MCS where $\text{speedup}=\frac{\text{MCS}_\text{runtime}}{\text{RMSS}_\text{runtime}}$. We ran all simulations on a Linux system with an Intel processor and a 32-core CPU. Parallelized MCS was solved for 10,000 samples using a minimum of 10 and up to 32 cores.

\input{sections/discussion.tex}

%% file: sections/discussion.tex
\subsection{Discussion}\label{discussion}
The accuracy and computational performance results illustrate the strengths and weaknesses of RMSS.

\subsubsection{Accuracy}

\begin{figure}[h!t]
  {
    \begin{center}
      \includegraphics[trim=5 5 5 0,clip, width=0.65\columnwidth]{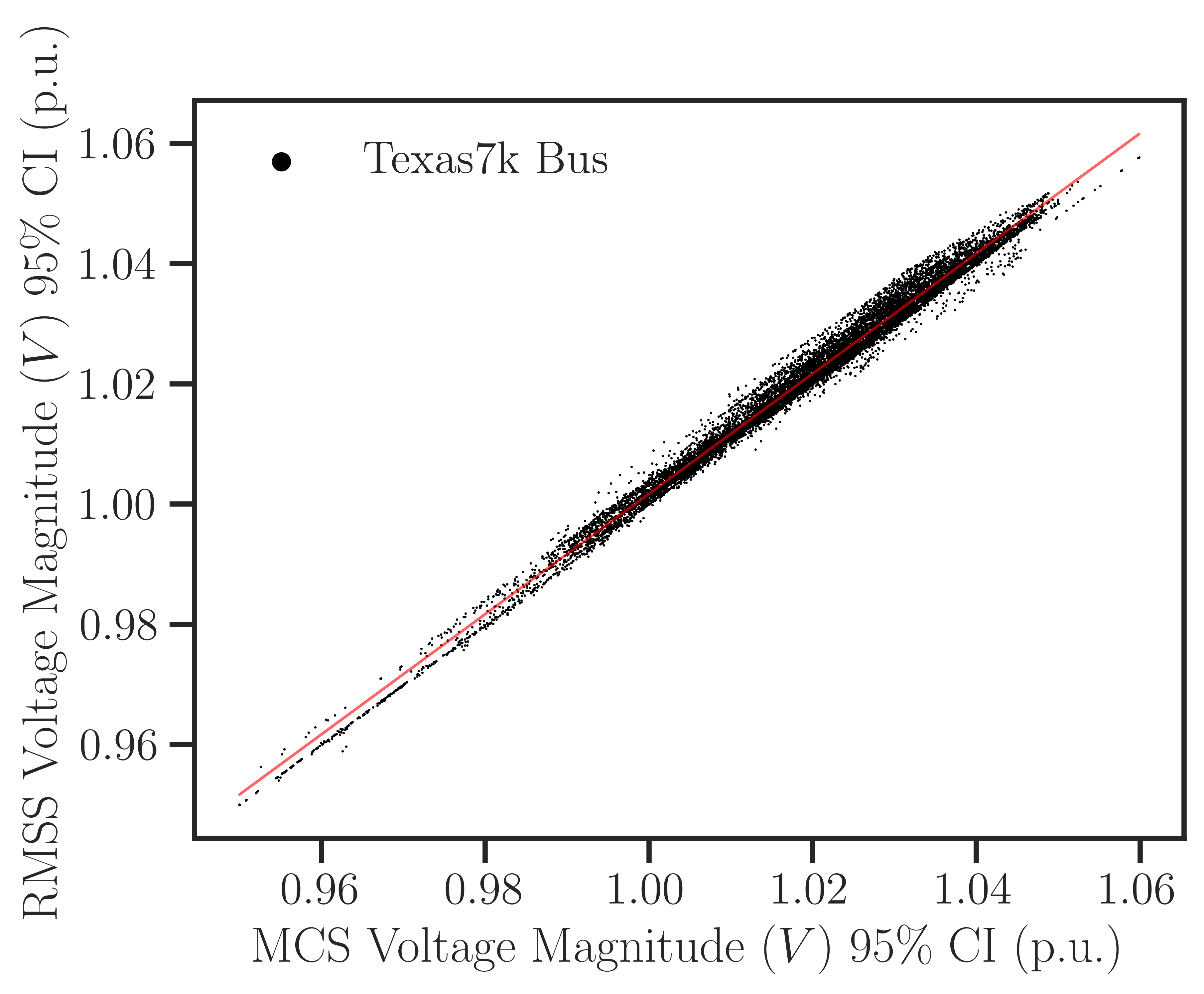}
    \end{center}
  }
    \caption{A strong, positive, and linear association exists between MCS $V$ 95\% CI and RMSS $V$ 95\% CI with relatively few outliers. The results are from a Texas7k low wind test system with 10,343 \stochps.}
    
    \label{fig:vmag-results}
\end{figure}

The bus voltage magnitude ($V$) results from a low wind scenario are in Fig. \ref{fig:vmag-results}. There are 10,343 \stochps{} in this scenario. The \stochps{} are the $P$ of renewable generation and the $P$ and $Q$ of all loads in the network. In the plot, we see a strong, positive, and linear association between MCS $V$ 95\% CI and RMSS $V$ 95\% CI with relatively few outliers. In all Texas7k low wind day test systems we've examined, we've found that there is always a strong linear relationship between MCS $V$ 95\% CI and RMSS $V$ 95\% CI.

\begin{figure}[h!t]
  {
    \begin{center}
      \includegraphics[trim=5 5 5 0,clip, width=0.65\columnwidth]{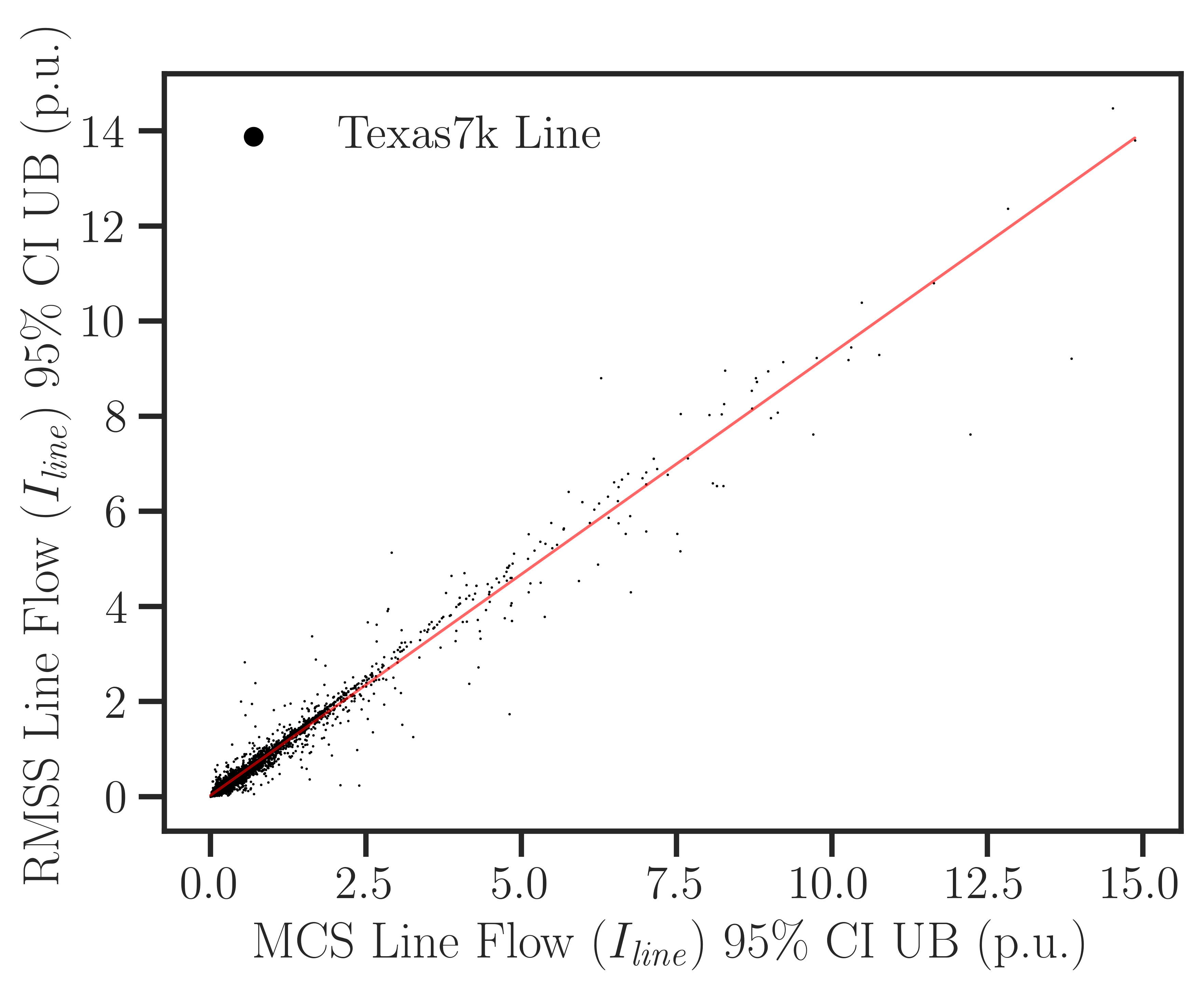}
    \end{center}
  }
    \caption{A moderately strong, positive, and linear association exists between MCS $\Iline$ 95\% CI upper bound (UB) and RMSS $\Iline$ 95\% CI upper bound (UB). The results are from a Texas7k low wind test system with 10,343 \stochps.}
    
    \label{fig:line-flow-results}
\end{figure}

Fig. \ref{fig:line-flow-results} shows current magnitude line flow ($\Iline$) results from a low wind scenario where the \stochps{} are the $P$ of renewable generation and the $P$ and $Q$ of all loads in the network. There is a moderately strong, positive, and linear association between MCS $\Iline$ 95\% CI upper bound (UB) and RMSS 95\% CI upper bound (UB). We've found similar results for all $\Iline$ in other Texas7k low wind day test systems studied.

In comparison to Fig. \ref{fig:vmag-results}, we observe more outliers in Fig. \ref{fig:line-flow-results}. There are more outliers for $\Iline$ because RMSS presently uses a single deterministic power flow for the mean \crits. When $V$ is the \crit{}, the assumption that a single deterministic power flow can describe the nominal performance is empirically found to be accurate. However, in the case of the line flows, it is possible that the mean \crit{} obtained from a single power flow is different from the MCS mean \crit. Hence, the worst-case predictions for $\Iline$ can be less accurate since their nominal values are not exact. In case a more accurate match to MCS tail regions is required for analysis, it is possible to obtain a more representative estimate for any mean \crit{} using second-order sensitivities \cite{cacuci-2018} at a higher computational cost.

\subsubsection{Computational Performance}
\begin{figure}[h!t]
  {
    \begin{center}
      \includegraphics[trim=5 5 5 0,clip, width=0.65\columnwidth]{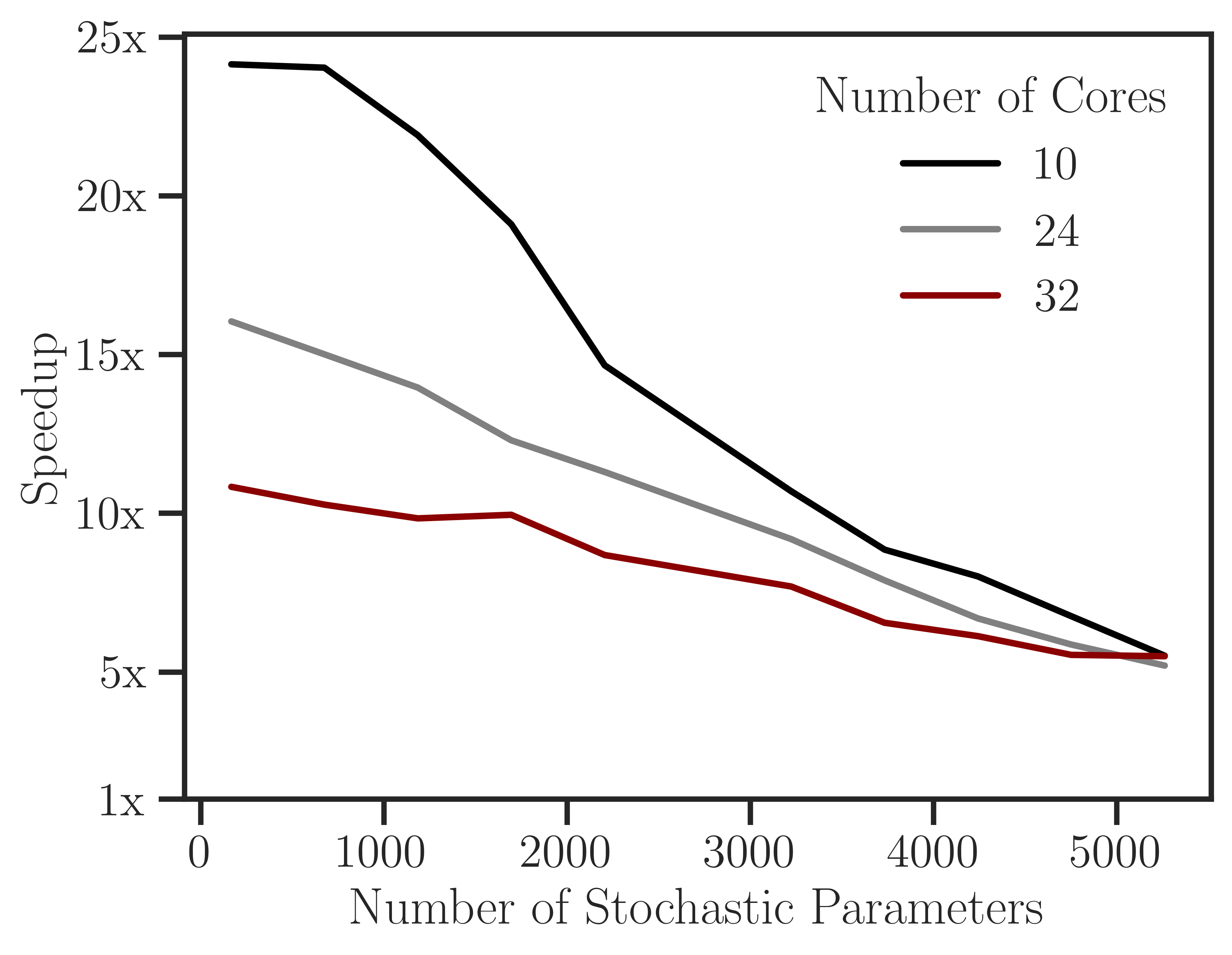}
    \end{center}
  }
    \caption{A comparison of the computational performance of RMSS with MCS for a Texas7k low wind day test system. Both RMSS and MCS are implemented using parallel processing.}
    \label{fig:perf-results}
\end{figure}

RMSS is compared with MCS, as shown in Fig. 8, with the speedup indicating the improvement with RMSS for the Texas7k low wind system. RMSS and MCS are both implemented using parallel processing on the same machine, but MCS is embarrassingly parallel, and runtime scales more effectively for it with the number of cores. As shown in Fig. \ref{fig:perf-results}, the runtime speedup of RMSS over MCS is 24x for 10 cores but drops to 16x for 24 cores and 11x for 32 cores. Notice that while this clearly demonstrates the benefit of parallel processing for MCS, there is a compression of scaling benefit with the number of cores.

Perhaps more importantly, it should be noted that the runtime per MCS power flow sample is nonuniform, and some ill-conditioned network scenarios that are hard to solve using power flow can dramatically increase the runtime per MCS sample. 

Additionally, some networks may have a significant portion (e.g., 10\%) of MCS samples as failed (infeasible: unable to satisfy AC network constraints). In this case, any failed sample will likely lie in the tail of a distribution. As such, there would be no way of quantifying the quality of solutions and estimating the true distributions of \crits{} with MCS. Under the scenario of a high number of failed samples, a decision maker might make a less informed decision based on the result, given the inaccuracy of the distribution estimate. RMSS overcomes this challenge since it does not require repeated samples.

If generalizable, the computational performance advantage of RMSS over MCS could indicate that RMSS can help analysts make quick decisions during emergencies or perform probabilistic contingency analyses. The speed improvement of RMSS means that grid analysts do not have to perform MCS for millions of samples and hope all MCS power flow samples converge to a feasible solution.
 
%% file: sections/conclusion.tex
\section{Conclusion}\label{conclusion}This work presents an alternate probabilistic load flow method known as \RMSS{} (RMSS). Our method predicts \crit{} violations and estimates the \stochps{} that will result in \crit{} violations. Results show that for the Texas7k low wind test system analyzed, (1) RMSS is moderately accurate with a reasonable estimate for mean \crits{}, and (2) RMSS is up to 24x faster in runtime speed than MCS implemented with parallel processing. In our future work, we will develop other methods for estimating mean critical performance, handling nonnormal RVs, improving computational performance, and mitigating worst-case \crit{} violations by designing an optimization formulation. Based on the evidence, we believe RMSS could be very useful to grid analysts, especially during day-ahead operational planning and when quick decisions are needed.